# Flexible generation of structured terahertz fields via programmable exchange-biased spintronic emitters

Shunjia Wang[1†], Wentao Qin[1†], Tongyang Guan[1†], Jingyu Liu[2†], Qingnan Cai[1], Sheng Zhang[1], Lei Zhou[1], Yan Zhang[2*], Yizheng Wu[1*], Zhensheng Tao[1*]

[1]State Key Laboratory of Surface Physics and Department of Physics and Key Laboratory of Micro and Nano Photonic Structures (MOE), Fudan University, Shanghai 200433, China.

[2]Beijing Key Laboratory of Metamaterials and Devices, Key Laboratory of Terahertz Optoelectronics, Ministry of Education, Beijing Advanced Innovation Center for Imaging Theory and Technology, Department of Physics, Capital Normal University, Beijing 100048, China.

[†]These authors contributed equally to this work.

[*]Emails: yzhang@cnu.edu.cn; wuyizheng@fudan.edu.cn; zhenshengtao@fudan.edu.cn.

**Abstract**

Structured light, particularly in the terahertz frequency range, holds considerable potential for a diverse range of applications. However, the generation and control of structured terahertz radiation pose major challenges. In this work, we demonstrate a novel programmable spintronic emitter that can flexibly generate a variety of structured terahertz waves. This is achieved through the precise and high-resolution programming of the magnetization pattern on the emitter's surface, utilizing laser-assisted local field cooling of an exchange-biased ferromagnetic heterostructure. Moreover, we outline a generic design strategy for realizing specific complex structured terahertz fields in the far field. Our device successfully demonstrates the generation of terahertz waves with diverse structured polarization states, including spatially separated circular polarizations, azimuthal or radial polarization states, and a full Poincaré beam. This innovation opens a new avenue for designing and generating structured terahertz radiations, with potential applications in terahertz microscopy, communication, quantum information, and light-matter interactions.

**Introduction**

Structured light, referring to custom light fields with tailored intensity, polarization and phase[1,2], is essential in various fields of research and applications, including imaging[3], microscopy[4,5], communication[6], quantum information[7], and light-matter interactions[8,9]. The terahertz frequency range is particularly intriguing due to its nonionizing nature, excellent penetration capability, and its unique ability to carry fingerprint signatures of condensed matter and biomolecules[10]. Thus, the expertise to generate and control structured terahertz radiation holds great promise for advancing terahertz applications. For instance, precise structuring of terahertz wavefronts could enhance the spatial resolution of terahertz microscopy[11,12], enable terahertz holographic imaging[13], and facilitate novel techniques for terahertz spectroscopy and sensing[14]. Moreover, structured terahertz waves with orbital angular momentum (OAM) and OAM multiplexing could potentially expand terahertz communication bandwidth[15,16].

Spatial light modulators based on liquid crystals and digital micromirror devices are commonly used to generate structured light at optical frequencies[17,18]. However, their applications at terahertz frequencies present challenges due to the necessity for substantially larger sizes of liquid crystals and digital micromirrors. The increased sizes result in higher losses and voltage requirements for liquid crystals, as well as manufacturing challenges for micromirrors. On the other hand, metasurfaces present an alternative technology for enhanced interaction with terahertz waves, where their subwavelength meta-atoms contribute to higher spatial resolution and larger modulation depth[19–22]. Nevertheless, these advantages often come with a limitation on the working bandwidth due to resonance effects. Furthermore, most terahertz metasurfaces demonstrated thus far are passive and unifunctional, often requiring the design and manufacture of a new metasurface device for each specific wave-manipulation function. Though a few reconfigurable metasurfaces have been realized in the terahertz regime using novel materials such as semiconductors[23,24], phase-transition materials[25,26], two-dimensional materials[27,28] and nonlinear crystals[29], they still encounter limitations including poor scalability, narrow bandwidth, and limited functionality.

Recent advancements in spintronic terahertz emitters present a new avenue to efficiently generate and actively control terahertz waves within a single device. These spintronic emitters are heterostructure devices, comprising nanometer-thick ferromagnetic (FM) and heavy metal (HM) layers. In such a device, one can effectively generate and control ultrabroadband terahertz emission via the spin-to-charge conversion process through the inverse spin Hall effect (ISHE)[30–32], occurring within the HM layer with strong spin-orbit coupling. Furthermore, by fabricating the metallic thin films into subwavelength structures (meta-atoms), spintronic-metasurface emitters can produce terahertz radiations with circular polarizations[33], controllable wavefronts[34], and OAM[35].

Nevertheless, in these configurations, the external field can only uniformly saturate the FM magnetization within each meta-atom, generating local terahertz waves with the same polarization. This constraint significantly limits the flexibility and diversity for structured terahertz wave generation. In particular, the current approach hinders simultaneous manipulation of the spin and orbital angular momenta. To realize diverse and complex functionalities, individual metasurface devices still need to be manufactured. Although previous research has employed an inhomogeneous field distribution to overcome this limitation and generate elliptically polarized terahertz fields[36], this approach lacks the necessary spatial resolution and field complexity for arbitrary terahertz-wave manipulation.

In this work, we demonstrate a novel programmable spintronic emitter with remarkable capability to generate a variety of structured terahertz waves, with simultaneous control over both spin and orbital angular momenta. This novel capability is enabled by the flexible, micrometer-resolution programming of the magnetization pattern at the emitter's surface, which arises from the unidirectional magnetic anisotropy within an FM layer due to its exchange-coupling with an adjacent antiferromagnetic (AFM) layer[37]. The local magnetization direction hence becomes flexibly programmable through a local field cooling process. Consequently, with high degrees of freedom, we can meticulously design and generate broadband terahertz beams featuring diverse structured polarization states. These include spatially separated circular polarizations, azimuthal or radial polarization states, and the creation of a full Poincaré beam, showcasing the immense potential of our programmable spintronic emitter.

**2. Programming of exchange-biased spintronic emitters**

Our experiments are based on a trilayer heterostructure, which is composed of a 6-nm-thick IrMn (AFM) layer, a 3-nm-thick $Fe_{21}Ni_{79}$ (FM) layer, and a 2-nm-thick Pt (HM) layer, as illustrated in Fig. 1a. The entire heterostructure is deposited on a 0.5-mm-thick, double-side polished $Al_2O_3(0001)$ substrate. Further details about sample preparation can be found in Methods. The interface exchange interaction between the FM and IrMn layers induces a strong unidirectional magnetic anisotropy within the FM layer, resulting in a net magnetization (**M**) in the absence of an external magnetic field[37].

Under femtosecond laser excitation, nonequilibrium spin-polarized currents (**j**$_s$) transport from the FM to the Pt (HM) layer, with the spin polarization of **j**$_s$ aligning with **M**. The strong spin-orbit coupling of HM then converts the longitudinally propagating **j**$_s$ to transverse charge currents **j**$_c$, according to the ISHE[31]: $\mathbf{j}_c = \gamma \mathbf{j}_s \times \mathbf{M}/|\mathbf{M}|$, where $\gamma$ represents the spin-Hall angle of HM. As a result, the trilayer heterostructure can emit ultrashort terahertz pulses in a field-free environment upon excitation by femtosecond laser pulses[38], with the polarization of the terahertz field perpendicular to **M**.

The programming of the magnetization pattern, **M**(**r**), of the spintronic emitter is accomplished through the laser-assisted magnetic programming (LAMP) method[39], as illustrated in Figs. 1b-d. Initially, a uniform unidirectional magnetic anisotropy is established within the FM layer during the sample growth process (see Methods), causing the hysteresis curves to shift due to an exchange bias (Fig. 1b).

In the subsequent step (Fig. 1c), a femtosecond laser beam (referred to as the writing beam) with a power of $P_w$ is focused onto the designated location in the presence of a writing magnetic field **H**$_w$=1000 Oe. The writing laser beam possesses a pulse duration of 110 fs, a repetition rate of 36 MHz, and a central wavelength of 1030 nm. The localized laser heating raises the temperature of the heterostructure beyond the Néel temperature[37], $T_N$, while ensuring it remains below the Curie temperature of the FM layer. As the laser spot traverses along the writing direction with a velocity of $v_w$, the previously heated area of the AFM film undergoes a local field cooling. This field-cooling process induces a rearrangement of the AFM spins within the patterned region due to the exchange coupling with the adjacent FM layer when the spin polarization of the FM layer is uniformly aligned in the direction of $\mathbf{H}_w$ (Fig. 1c).

Consequently, once the writing laser $P_w$ and writing field $\mathbf{H}_w$ are both removed (Fig. 1d), the magnetization within the FM layer in the programmed area (orange arrows) is pinned by the new unidirectional magnetic anisotropy in the direction of $\mathbf{H}_w$, while the magnetization in the unprogrammed area (white arrows) returns to its original state. The exchange bias in the patterned region is thus modified by this local field cooling process. Further details about the LAMP method can be found in Methods and Supplementary Materials (SM) section S1.

In Fig. 2a and b, we demonstrate the flexible control of exchange bias through the LAMP method. In these experiments, we utilize the same femtosecond laser to excite the specific regions of the heterostructure where the magnetization programming was performed, ensuring that the excitation laser power remained below the threshold power (0.46 W, see below). The hysteresis curves are obtained by plotting the signed amplitudes of emitted terahertz waves (Figs. 2c and d) against the magnetic field strength. The hysteresis curves before programming are plotted as the dashed lines, which exhibits a robust exchange-bias shift of $H_{EB}\approx500$ Oe along $y$ (Fig. 2a), and a typical hard-axis-loop behavior along $x$ (Fig. 2b).

In Fig. 2a, we explore two scenarios during the LAMP process: applying $\mathbf{H}_w$ along $+y$ (I) and $-y$ (II) directions. Note that under these conditions, the polarization of terahertz radiation aligns with the $x$-axis (Fig. 2c). Here, we observe a clear shift of the exchange bias when $\mathbf{H}_w$ is applied in the opposite directions along $y$, leading to the opposite polarities of the emitted terahertz waves under zero field (Fig. 2c). Similarly, in Fig. 2b, we further demonstrate control of the exchange bias along the +*x* (III) and -*x* (IV) directions, with the corresponding terahertz waveforms shown in Fig. 2d. It is important to note that the temporal terahertz waveforms exhibit remarkable similarity (Figs. 2c and d), which is essential for the broadband application of our method.

In Fig. 2e, we present the Kerr microscopy images of the programmed magnetic domains under different writing laser power $P_w$. The diameter of the writing laser beam ($D_0$) is approximately 80 μm on the sample surface, moving at a linear speed of $v_w$=500 μm s$^{-1}$ (see Methods). We observe that the domain width can be precisely controlled by $P_w$. To quantify this effect, we plot the magnetic domain width ($D_w$) as a function of $P_w$ in Fig. 2f. The data can be well fitted to an empirical model $D_W = \frac{D_0}{\sqrt{2}}\sqrt{-\ln\left(\frac{P_{th}}{P_w}\right)}$, where $P_{th}$=0.46 W represents a threshold power (see SM Section S1). This threshold power ($P_{th}$) denotes the minimum laser power required to reach $T_N$ under the current laser conditions. Furthermore, our results reveal that the heterostructure suffers permanent damage when $P_w$ exceeds 1.5 W (see SM Fig. S2). For subsequent demonstrations, we prudently select a writing power of $P_w$=1.2 W, providing a spatial resolution for the magnetization-pattern programming of approximately 40 μm.

**3. Design strategy of magnetization patterns**

In order to generate a target far-field terahertz beam with complex spatial and polarization characteristics, it is important to establish a generic strategy for designing the magnetization pattern **M(r)**, where **r** denotes the position vector in the transverse plane (Fig. 1e). Here, we adopt the paraxial approximation and assume that the terahertz electric fields are confined in the transverse plane. The target far-field terahertz beam, under the circular polarization basis ($\mathbf{e}_L$ and $\mathbf{e}_R$), can be described as

$$\mathbf{E}_{tar}(\mathbf{r}) = A_{tar}^L(\mathbf{r})e^{i\Phi_L(\mathbf{r})}\mathbf{e}_L + A_{tar}^R(\mathbf{r})e^{i\Phi_R(\mathbf{r})}\mathbf{e}_R. \quad (1)$$

In Eq. (1), $A_{tar}^{L(R)}$ and $\Phi_{L(R)}$ denote the far-field amplitude and phase of the left- (right-) handed circular polarization [LCP (RCP)] component, respectively. The local near-field (NF) terahertz waves at the emitter's surface, denoted as $\mathbf{E}_{NF}(\mathbf{r})$, can then be calculated following scalar diffraction theory (Fig. 1e):

$$\mathbf{E}_{NF}(\mathbf{r}) = \int \tilde{\mathbf{E}}_{tar}(\mathbf{k}_\parallel) G(\mathbf{k}_\parallel) e^{i\mathbf{k}_\parallel \cdot \mathbf{r}} \, d\mathbf{k}_\parallel. \quad (2)$$

Here, $\tilde{\mathbf{E}}_{tar}(\mathbf{k}_\parallel) = \int \mathbf{E}_{tar}(\mathbf{r}) e^{-i\mathbf{k}_\parallel \cdot \mathbf{r}} d\mathbf{r}$ denotes the Fourier-space representation of the target field, with $\mathbf{k}_\parallel$ being the transverse wavevector (Fig. 1e). The transfer function of free-space propagation, $G(\mathbf{k}_\parallel)$, is given by

$$G(\mathbf{k}_\parallel) = \begin{cases} e^{iz_0\sqrt{|\mathbf{K}_0|^2 - |\mathbf{k}_\parallel|^2}}, & |\mathbf{k}_\parallel| < |\mathbf{K}_0| \\ 0, & |\mathbf{k}_\parallel| \geq |\mathbf{K}_0| \end{cases}. \quad (3)$$

Here, $\mathbf{K}_0$ represents the terahertz wavevector with $|\mathbf{K}_0| = \frac{2\pi}{\lambda}$ (Fig. 1e), where $\lambda$ is the terahertz wavelength, and $z_0$ is the distance from the emitter's surface to the transverse plane of interest.

To generate complex structured terahertz fields in the far field, it is necessary to exert control over both the amplitude and polarization states of $\mathbf{E}_{NF}(\mathbf{r})$. First, the amplitude $|\mathbf{E}_{NF}(\mathbf{r})|$ can be regulated by varying the local intensity of the excitation laser. In this work, as lasers with Gaussian profiles were employed to excite various programmed emitters, the amplitude design on $\mathbf{E}_{NF}(\mathbf{r})$ was not pursued. Second, the linear polarization of $\mathbf{E}_{NF}(\mathbf{r})$ aligns with the flowing direction of the transverse current density $\mathbf{j}_c(\mathbf{r})$ at the same location following Ohm's law: $\mathbf{E}_{NF}(\mathbf{r}) \propto \mathbf{j}_c(\mathbf{r})$. The orientation of the local magnetization $\mathbf{M}(\mathbf{r})$ is then derived by $\frac{\mathbf{M}(\mathbf{r})}{|\mathbf{M}(\mathbf{r})|} = \frac{\mathbf{j}_c(\mathbf{r}) \times \mathbf{j}_s}{|\mathbf{j}_c(\mathbf{r}) \times \mathbf{j}_s|}$, according to the ISHE[31]. Further details about the design strategy can be found in Methods.

## 4. Generation of complex structured terahertz fields

Following the design strategy, we first demonstrate the generation of terahertz beams with spatially separated circular polarizations. Here, the target terahertz field can be expressed as

$$\mathbf{E}_{tar}(x,y) = A_0 e^{-\left[\frac{(x+x_0)^2+y^2}{w^2}\right]} e^{-ik_0 x} \mathbf{e}_L + A_0 e^{-\left[\frac{(x-x_0)^2+y^2}{w^2}\right]} e^{ik_0 x} \mathbf{e}_R, \quad (4)$$

where $A_0$ is the field amplitude, $w$ is the Gaussian beam radius, and $k_0$ is the transverse wavevector defined by $k_0 = \frac{2\pi}{\lambda}|\sin\varphi_e|$ with $\varphi_e$ being the emission angle. The transverse offset of the beams at the target plane, denoted as $x_0$, is given by $x_0 = z_0 |\tan\varphi_e|$. For this demonstration, we intentionally set $\varphi_e = \pm 30°$ for the terahertz waves with a wavelength of $\lambda = 300$ μm (corresponding to 1 THz). The radius of the excitation laser beam is set to $w = 1.8$ mm.

In Fig. 3a and b, we respectively plot the local terahertz NF $\mathbf{E}_{NF}(\mathbf{r})$ and the corresponding magnetization pattern $\mathbf{M}(\mathbf{r})$ obtained from our design strategy (see SM Section S3). The separation of the terahertz waves carrying opposite spin angular momenta results in a rotating local-field polarization along $x$, and its spatial periodicity $\Lambda$ is 600 μm.

For the experimental realization, we program a heterostructure sample with transverse dimensions of 1 cm×1 cm, following the pattern of $\mathbf{M}(\mathbf{r})$ shown in Fig. 3b. Since the polarization of the emitted terahertz wave is perpendicular to the local magnetization, we employ a focused-beam excitation setup to evaluate the accuracy of the programmed heterostructure. In this setup, a laser beam is focused to excite the local magnetic domains, and the laser spot is scanned along the $x$-axis following the trajectory shown in Fig. 3a. Simultaneously, the amplitudes of $E_x$ and $E_y$ terahertz components are recorded (see Methods and SM Section S2). The results, as depicted in Fig. 3c, exhibit sinusoidal oscillations as a function of the scanning position ($x_s$) with a phase offset of $\pi/2$ between $E_y$ and $E_x$. The obtained results are in excellent agreement with our wave-propagation simulations (see Methods), which exhibits a chiral variation of $\mathbf{M}(\mathbf{r})$ along $x$ with a superlattice constant of $\Lambda=600$ μm. These results clearly demonstrate the successful programming of the exchange-biased emitter.

The emission of terahertz radiation from the entire programmed heterostructure is excited using an amplifier laser beam (25 fs pulse duration, 10 kHz repetition rate, and a central wavelength of 1030 nm) with a large beam radius of approximately 1.8 mm, effectively covering multiple lattice constants. The spatial distribution of the polarization states is subsequently detected using an angle-resolved detection setup (see Methods and SM Section S7).

In Figs. 3d and e, we respectively present the experimental spatial distribution of the LCP and RCP terahertz radiations. The results reveal that the LCP and RCP radiations are emitted at different angles, with LCP observed at $\varphi_e<0$ and RCP at $\varphi_e>0$. Moreover, the terahertz waves with the frequency of 1 THz are emitted at $\varphi_e=\pm30°$, consistent with our design. The diffraction from the superlattice further results in the spatial chirp of the terahertz waves at different $\varphi_e$, as described by $f = \frac{c}{\Lambda|\sin\varphi_e|}$ (Figs. 3d and e), where $f$ represents the terahertz frequency and $c$ is the speed of light. These results can be well reproduced by our wave-propagation simulations (Fig. 3f and g), where the finite spot size of the excitation laser and the experimental terahertz spectrum are considered (see Methods). In Figs. 3h and i, we present the experimental temporal profiles of terahertz radiations emitted at $\varphi_e$=-30º and 30º, respectively, clearly showing distinct LCP and RCP chirality.

In the second case, we demonstrate the generation of an azimuthally polarized terahertz beam. The target terahertz field can be expressed under the radial coordinate ($r$, $\phi$) by

$$\mathbf{E}_{tar}(r,\phi) = A_0 e^{-\frac{r^2}{w^2}} e^{i\left(\phi+\frac{\pi}{2}\right)} \mathbf{e}_L + A_0 e^{-\frac{r^2}{w^2}} e^{-i\left(\phi+\frac{\pi}{2}\right)} \mathbf{e}_R. \quad (5)$$

In Fig. 4a and b, we plot the local NF $\mathbf{E}_{NF}(\mathbf{r})$ and the magnetization pattern $\mathbf{M}(\mathbf{r})$, respectively. The design process is described in SM Section S4. The generation of azimuthally polarized terahertz fields requires a radially polarized distribution of $\mathbf{M}(\mathbf{r})$ (Fig. 4b).

For the experimental realization, we reprogram the same heterostructure according to the pattern of $\mathbf{M}(\mathbf{r})$ in Fig. 4b. In the focused-beam excitation setup, we scan the laser spot along a circular trajectory around the center, with a radius of $r_s$=250 μm (Fig. 4a). The $E_x$ and $E_y$ amplitudes are plotted in Fig. 4c, exhibiting sinusoidal behaviors against the scanning angle $\theta_s$ with a phase offset of π/2. These results confirm the radial distribution of $\mathbf{M}(\mathbf{r})$.

The emission of terahertz radiation from the entire heterostructure is excited using a femtosecond laser amplifier (50 fs pulse duration, 1 kHz repetition rate, and a central wavelength of 780 nm). The excitation beam is normal incident to the emitter's surface with a beam radius of 4.5 mm, its beam center aligned with the center of the programmed structure. The spatio-temporal profiles of the radiation are measured using a polarization-resolved terahertz digital holographic imaging system[40] (TDHIS, see Methods and SM Section S8).

In Figs. 4d1 and d2, we present the experimentally measured spatial profiles of the intensity ($|\mathbf{E}_{tar}|^2$) and phase of the terahertz radiation, respectively, for the polarization component parallel to $y$ ($\parallel y$), while Figs. 4d3 and d4 depict the corresponding results for the polarization orthogonal to $y$ ($\perp y$). The observed profiles exhibit two distinct lobes along a line perpendicular to the polarization direction, with the polarities of the terahertz waves flipping across the beam center. These results provide clear evidence for the generation of azimuthally polarized terahertz beams. Our wave-propagation simulations reproduce these experimental results, as shown in Figs. 4e1-e4.

Interestingly, despite employing a Gaussian beam profile for the target terahertz field [see Eq. (5)], the rapid polarization rotation within a small area at the beam center results in diffraction with a transverse wavevector exceeding $2\pi/\lambda$, which is forbidden in free-space wave propagation [see Eq. (3)]. As a result, both the experimental and simulation results exhibit a suppression of the field intensity at the beam center.

In Figs. 4d-e, we present the results obtained specifically at the terahertz frequency of $f$=0.47 THz. The results at other terahertz frequencies are provided in SM Section S4, highlighting the broad bandwidth of the generated azimuthally polarized terahertz beams. Additionally, by programming $\mathbf{M}(\mathbf{r})$ to an azimuthally polarized distribution, we can effectively generate broadband radially polarized terahertz radiation (see SM Section S5).

By achieving flexible and subwavelength control over the polarization states of local terahertz emitters, we acquire the capability to manipulate both the spin and orbital angular momenta of the far-field terahertz beams. In the last case, we demonstrate this capability by generating a terahertz beam that provides all the states of polarization in the beam cross-section, also known as the full Poincaré (FP) beam[41].

The target terahertz beam is a superposition of two lowest order Laguerre-Gaussian (LG) beams (with a radial index of $p$=0), carrying opposite circular polarizations. Here, we choose the scalar LG beam (the Gaussian beam) for the LCP field, while assigning the LG beam with OAM of $l$=-2 to the RCP field. Consequently, the target field can be defined as

$$\mathbf{E}_{tar}(r,\phi) = A_R\, C_{lp}^{LG}\left(\frac{r\sqrt{2}}{w_R}\right) e^{-\frac{r^2}{w_R^2}} L_p^{|l|}\left(\frac{2r^2}{w_R^2}\right) e^{il\phi}\, \mathbf{e}_R + A_L e^{-\frac{r^2}{w_L^2}} e^{i\vartheta}\mathbf{e}_L, \qquad (6)$$

where $A_{R(L)}$ and $w_{R(L)}$ represent the field amplitude and the beam radius of the RCP (LCP) light, respectively, $\vartheta$ denotes a phase difference between the RCP and LCP components, $L_p^{|l|}$ is the generalized Laguerre polynomials, and $C_{lp}^{LG}$ is the normalized constant: $C_{lp}^{LG} = \sqrt{\frac{2p!}{\pi(p+|l|)!}}$. In this design, we set $w_R$=550 μm and $w_L$=333 μm, respectively, to fit the LCP beam within the singularity at the center of the RCP beam. The target terahertz frequency is $f$=0.5 THz, and the phase $\vartheta$ is set to be -1.2π.

Figures 5a and b display the distributions of $\mathbf{E}_{NF}(\mathbf{r})$ and $\mathbf{M}(\mathbf{r})$, respectively, obtained from our design strategy (see SM Section S6). The spin angular momentum of the emitted terahertz wave is generated by the spatial rotation of $\mathbf{M}(\mathbf{r})$, while the vortex Fresnel zone plate pattern of $\mathbf{M}(\mathbf{r})$ leads to the generation and focusing of the OAM beam. The accuracy of our programmed heterostructure is again verified by the focused-beam excitation measurements (Fig. 5c). A complete mapping of the entire heterostructure is provided in SM Fig. S12.

The terahertz emission is excited by the same femtosecond laser amplifier used in the last case and measured by the TDHIS system. The excitation beam radius is 4.5 mm. In the right panel of Fig. 5d, we plot the experimental intensity profiles and phases of the RCP and LCP terahertz components at a frequency of $f$=0.5 THz. We observe that due to the aberration of the vortex zone plate, the vortex at the center of the RCP beam splits into two in the far field. Still, the phase of the RCP beam varies by $2\pi$ as it rotates by an angle $\phi$ of -180º around the two vortices, demonstrating the generation of OAM with $l$=-2. The LCP beam exhibits a simple Gaussian profile with a flat phase distribution, consistent with our design. The results obtained by the wave-propagation simulation are also shown in the right panel of Fig. 5e for comparison.

We calculate the Stokes parameters $S_0$, $S_1$, $S_2$ and $S_3$, allowing us to further determine the orientation and ellipticity of the terahertz fields across the beam cross-section. In Fig. 5d, we plot the experimental distribution of polarization ellipses. The interference of $l$=0 and $l$=-2 vortex beams results in a complex polarization distribution, exhibiting variations from LCP at the beam center to RCP at the fringes, and linear polarizations with changing orientations at the boundaries. We find that the polarization states cover the entire Poincaré sphere for a beam radius of 800 μm, confirming the generation of FP beam (see SM Section S6). The corresponding simulation results are shown in Fig. 5e.

**5. Discussion**

In this work, we show that the flexible programming of the exchange-biased magnetic heterostructure enables the direct generation of various structured terahertz beams with complex polarization distributions. In the above demonstrations, we did not perform amplitude design on $\mathbf{E}_{NF}(\mathbf{r})$, as lasers with Gaussian profiles were utilized to excite various programmed emitters. To exert control over local NF amplitudes, spatial light modulators can be further employed to manipulate the amplitude profiles of excitation lasers.

Furthermore, it is important to acknowledge that, owing to the inherent capability of generating only linearly polarized $\mathbf{E}_{NF}$ locally, a crucial constraint arises: the NF terahertz amplitudes for the LCP and RCP components must be equal at all locations, leading to $A_{NF}^L(\mathbf{r}) = A_{NF}^R(\mathbf{r})$ at the emitter's surface. As a consequence, both LCP and RCP terahertz fields are simultaneously generated in the far field. In situations where terahertz beams with a single type of circular polarization are of interest, such as the RCP beam with OAM of $l$=-2 in the last demonstration, different spatial phase gradients can be applied in the design on the LCP and RCP light beams, which allows for their spatial separation in the far field.

In conclusion, our study introduces a novel programmable spintronic emitter that enables the generation of diverse structured terahertz waves. This approach utilizes the laser-induced local field cooling of an exchange-biased magnetic heterostructure, which allows precise and micrometer-level programming of the magnetization pattern on the emitter's surface. Our method offers highly flexible and broadband control over the polarization states across the terahertz beam cross-section, which can open up new possibilities for various applications in terahertz microscopy, quantum information and communications.

**Methods**

**Sample fabrication.** The trilayer heterostructures Pt/Py/Ir$_{25}$Mn$_{75}$ were deposited on the double-side polished Al$_2$O$_3$(0001) substrates using magnetron sputtering, where Py is the permalloy Fe$_{21}$Ni$_{79}$. In our experiments, the Al$_2$O$_3$(0001) substrates were chosen because of their high thermal conductivity. We also prepared samples on the SiO$_2$ substrates, which exhibit a lower laser-damage threshold due to relatively lower thermal conductivity. All the Al$_2$O$_3$ substrates were first ultrasonically cleaned in acetone, alcohol, and deionized water. Then, they were annealed at 150°C for one hour in the sputtering chamber with a base pressure of $2 \times 10^{-8}$ torr. The substrates were mounted on a sample holder equipped with a pair of permanent magnets that can apply a bias field during sample growth, achieving a field strength of approximately 250 mT. All the Pt, Fe$_{21}$Ni$_{79}$, and Ir$_{25}$Mn$_{75}$ films were grown by sputtering at room temperature under a working Ar pressure of 3 mTorr. The Ir$_{25}$Mn$_{75}$ layer was grown through co-sputtering from the Mn and Ir$_{50}$Mn$_{50}$ targets. The thickness of all the layers was determined by the sputtering time, and their growth rates were calibrated by X-ray reflectivity measurement. The growth rate is 0.22 Å/s for Pt layers, 0.26 Å/s for the Fe$_{21}$Ni$_{79}$ layers, and 0.44 Å/s for the Ir$_{25}$Mn$_{75}$ layers. Finally, an 8-nm Al$_2$O$_3$ layer was deposited to prevent oxidation.

**LAMP method.** The LAMP method was performed using a custom-made magnetic writing system. The schematic is shown in SM Fig. S1. The writing laser used was a power-stabilized fiber oscillator (FIBRE SP, Nanjing Keyun Optoelectronics Ltd.) with a pulse duration of 110 fs, a repetition rate of 36 MHz, a central wavelength of 1030 nm and a maximum power of 3.3 W. The power of the writing laser beam was adjusted by a combination of a half waveplate and a polarizer. The writing laser beam was focused using a 20-cm lens onto the sample surface, resulting in a beam diameter of approximately 80 μm. The trilayer heterostructure was mounted on a 2D motion stage, allowing precise motion along the *x* and *y* directions with a precision of <1 μm and a full range of 25 mm. The magnetic patterns were written by the raster-scanning of the sample position at a linear speed of $v_w$=500 μm s$^{-1}$. For setting the angle of local magnetization, an orientated electromagnet was employed, providing a writing magnetic field (**H**$_w$) of 1000 oe in the *x-y* plane.

**Design strategy.** The local NF terahertz waves at the emitter's surface, **E**$_{NF}$(**r**), can be calculated following Eqs. (1-3). These local NF terahertz waves can be expressed in the circular polarization basis: $\mathbf{E}_{NF}(\mathbf{r}) = \sigma_{NF}^{L}(\mathbf{r})\mathbf{e}_L + \sigma_{NF}^{R}(\mathbf{r})\mathbf{e}_R$. The complex amplitude $\sigma_{NF}^{L(R)}(\mathbf{r})$ can be represented by $\sigma_{NF}^{L(R)}(\mathbf{r}) = A_{NF}^{L(R)}(\mathbf{r})e^{i\varphi_{L(R)}(\mathbf{r})}$, where $A_{NF}^{L(R)}$ and $\varphi_{L(R)}$ denote the amplitude and phase of the NF LCP (RCP) component, respectively.

Because only linearly polarized terahertz waves can be generated locally from the programmed magnetic domains, we have an important constrain: $A_{NF}^{L}(\mathbf{r}) = A_{NF}^{R}(\mathbf{r}) = A_{NF}(\mathbf{r})$. As a result, the NF terahertz waves can be expressed as $\mathbf{E}_{NF}(\mathbf{r}) = A_{NF}(\mathbf{r})[e^{i\varphi_L(\mathbf{r})}\mathbf{e}_L + e^{i\varphi_R(\mathbf{r})}\mathbf{e}_R]$.

In order to manipulate the terahertz NF, it is essential to control both the field amplitude, *A*$_{NF}$(**r**), and the phase terms $\varphi_{L(R)}$. The amplitude control can be achieved by manipulating the local intensity of excitation lasers, which can be realized by spatial light modulators. In this work, as lasers with Gaussian profiles were utilized to excite various programmed emitters, we did not perform the amplitude design on **E**$_{NF}$(**r**).

On the other hand, the phases of the LCP and RCP NF components $\varphi_{L(R)}$ can be controlled by the polarization orientation of local terahertz NF. To facilitate the design process, we further define

$$\begin{cases} \theta(\mathbf{r}) = [\varphi_R(\mathbf{r}) - \varphi_L(\mathbf{r})]/2 \\ \psi(\mathbf{r}) = [\varphi_R(\mathbf{r}) + \varphi_L(\mathbf{r})]/2 \end{cases}. \qquad (7)$$

Here, $\theta(\mathbf{r})$ represents the polarization orientation of $\mathbf{E}_{NF}(\mathbf{r})$ constrained in the range of [0, π), and $\psi(\mathbf{r})$ further defines the polarity of the terahertz NF. The polarization direction of $\mathbf{E}_{NF}(\mathbf{r})$ is then determined by $\Theta_{NF}(\mathbf{r}) = \begin{cases} \theta(\mathbf{r}), & \cos[\psi(\mathbf{r})] \geq 0 \\ \theta(\mathbf{r}) + \pi, & \cos[\psi(\mathbf{r})] < 0 \end{cases}$. Here, the continuous distribution of $\psi(\mathbf{r})$ is discretized to the regions where $\cos[\psi(\mathbf{r})] \geq 0$ and $\cos[\psi(\mathbf{r})] < 0$. When $\cos[\psi(\mathbf{r})] < 0$, it indicates a spatial reversal of local field polarization, representing a π phase shift of the local terahertz wave (see Fig. 1d).

In various demonstration cases, the pattern of $\theta(\mathbf{r})$ and $\psi(\mathbf{r})$ can be numerically derived from the target fields [Eqs. (4-6)] using the design strategy, thereby allowing for the calculation of the polarization distribution of $\mathbf{E}_{NF}(\mathbf{r})$. According to Ohm's law[31], the local charge current density is directly proportional to the local terahertz NF, given by $\mathbf{j}_c(\mathbf{r}) \propto \mathbf{E}_{NF}(\mathbf{r})$. Finally, the magnetization pattern is determined by the ISHE[31]: $\frac{\mathbf{M}(\mathbf{r})}{|\mathbf{M}(\mathbf{r})|} = \frac{\mathbf{j}_c(\mathbf{r}) \times \mathbf{j}_s}{|\mathbf{j}_c(\mathbf{r}) \times \mathbf{j}_s|}$.

**Wave-propagation simulations.** The simulation of far-field terahertz waves was carried out through wave-propagation simulations using a Python program. In this simulation, broadband terahertz pulses with a spectrum matching the experimental signal were assigned to emit from the surface of a programmed spintronic emitter. The polarization orientation and polarity of the locally emitted wave, $\mathbf{E}_{NF}(\mathbf{r})$, were set according to the magnetization pattern $\mathbf{M}(\mathbf{r})$. The amplitude of $\mathbf{E}_{NF}(\mathbf{r})$ was determined based on the local intensity of the excitation laser. The far-field spatio-spectral profiles of the $E_x$ and $E_y$ components were then obtained by considering the propagation of each frequency component of the terahertz waves to the far-field plane.

To simulate the experimental results obtained from the focused-beam excitation measurements, the simulations employed a small excitation laser spot transversing across the sample surface following the designated trajectories. The simulation results of the $E_x$ and $E_y$ fields were plotted in comparison with the experimental results.

**Focused-beam excitation setup.** The programmed magnetization pattern was verified using a focused-beam excitation setup. The schematic of the focused-beam excitation setup is shown in SM Fig. S4. This setup utilized the same fiber laser employed during the LAMP process. The laser beam was focused onto the programmed heterostructure achieving a beam radius of 250 μm, while the laser power density was limited to less than $9.2\times10^7$ W cm$^{-2}$. The emitted terahertz waves were collected and focused by a pair of off-axis parabolic mirrors. The $E_x$ and $E_y$ components were selected by a terahertz polarizer, and respectively detected using the electro-optic sampling (EOS) method. The EOS crystal utilized was a 1-mm thick GaP crystal. The heterostructure was moved by the 2D motion stage following the designed trajectory, and the terahertz fields emitted from the illuminated areas were recorded.

**Angle-resolved terahertz detection setup.** The large angle distribution of the terahertz waves with spatially separated circular polarizations was measured using an angle-resolved terahertz detection setup. The setup is illustrated in the SM Fig. S14. The excitation pulses were obtained from a Yb:KGW laser amplifier (PHAROS SP, Light-Conversion) with a repetition rate of 10 kHz, a center wavelength of 1030 nm. The spectrum of laser pulses was first broadened using an all-solid-state pulse compressor based on periodic layered Kerr medium (PLKM)[42,43]. The pulse compression from 170 fs to 25 fs was achieved through dispersion correction using a set of chirped mirrors. The excitation laser beam was focused on the sample surface, resulting in a beam radius of approximately 1.8 mm. The programmed sample was positioned on a rotation stage that allowed rotation about the *y*-axis, exposing terahertz radiations emitted at different angles ($\varphi_e$) to a polarization-resolved EOS detection setup. The excitation laser beam remained normal incident onto the sample surface. A terahertz polarizer was employed to select the $E_x$ and $E_y$ components.

**TDHIS setup.** The schematic of the setup is shown in SM Fig. S15. The system employs a Ti:Sapphire regenerative amplifier (Spitfire Ace, Spectra-Physics) to generate an ultrashort pulse of 50 fs with a central wavelength of 780 nm, a 1 kHz repetition rate, and a 9 mm spot diameter. The laser beam is divided into two beams: the generation beam and the detection beam. The generation beam excites the sample to generate the terahertz waves, detected by a 2-mm thick ZnTe crystal. Between the sample and the detection crystal, a 500-μm thick silicon wafer is placed to block the remaining generation beam. Using a half waveplate (HWP) and a polarizer (P), we control the polarization state of the probe beam. The probe beam, reflected by a beam splitter (BS) into the detection crystal, carries terahertz information after modulation by the terahertz radiation. This modulated probe beam is captured by the detection unit, which consists of a quarter waveplate (QWP), a Wollaston prism (WP), two lenses, and a charge-coupled device (CCD). We capture two mutually orthogonal components of the probe beams for differential probing. By altering the optical path difference between the generation and probe beams, we obtain a series of terahertz images at different delay times. Through Fourier transform analysis of the time signal on each pixel, we can extract amplitude and phase information at different frequencies on corresponding pixels. Furthermore, the detected terahertz polarization state can be adjusted by changing the polarization of the detected beam[44].

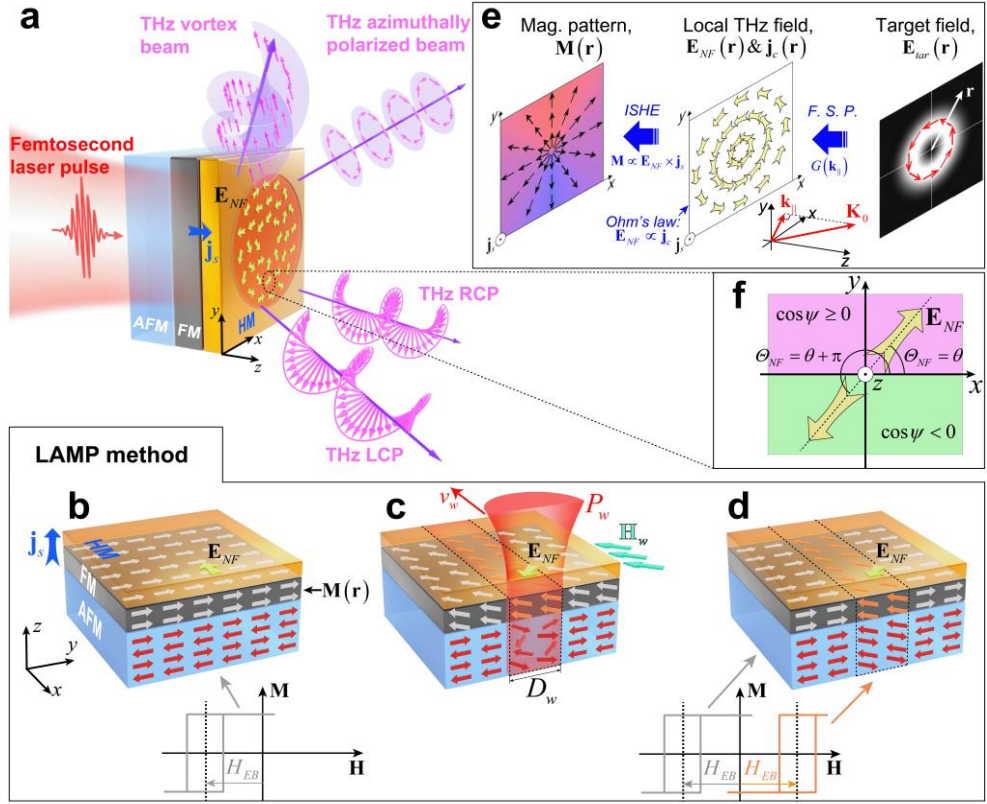

**Figure 1. Concept of the programmed exchange-biased spintronic terahertz emitter. a.** Schematic of the experimental setup. Femtosecond laser pulses are applied to excite the programmed exchange-biased spintronic emitter, which induces spin-polarized current ($\mathbf{j}_s$) and local terahertz NF $\mathbf{E}_{NF}$. Structured terahertz vector fields are generated in the far field. **b.-d.** Illustration of the LAMP process. **b.** The magnetization of the FM layer (white arrows) is uniformly pinned in one direction by the exchange interaction with the AFM layer (red arrows). The magnetic hysteresis loop before programming exhibits an exchange-bias field $H_{EB}$. **c.** The movement of the writing laser beam $P_w$ on the sample surface in the presence of the writing magnetic field $\mathbf{H}_w$ leads to a local field cooling, which resets the exchange bias direction in the programmed area (orange arrows) aligned with $\mathbf{H}_w$. **d.** When both $P_w$ and $\mathbf{H}_w$ are removed, the magnetization orientation in the FM layer is stabilized by the local exchange bias. The hysteresis loop in the programmed regions is modified. **e.** Illustration of the design strategy to derive the magnetization (mag.) pattern $\mathbf{M}(\mathbf{r})$ from the target field $\mathbf{E}_{tar}(\mathbf{r})$. F. S. P.: Free-space propagation. **f.** Illustration of different regions for the polarization angle of the local terahertz NF, $\Theta_{NF}$, determined by the design strategy.

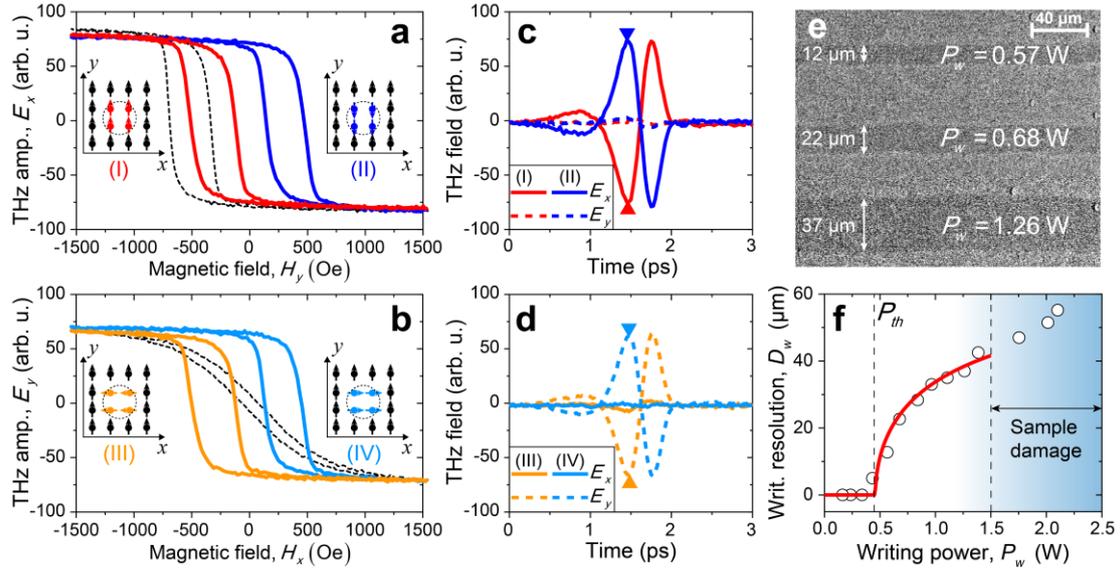

**Figure 2. Results of the LAMP process. a.** The hysteresis curves along the *y* axis, obtained by measuring the signed amplitude of $E_x$ as a function of the applied magnetic field $H_y$. The dashed lines are the results obtained from the unprogrammed regions. The red and blue solid lines represent the results from the programmed areas with $\mathbf{H}_w$ along +*y* (I) and along -*y* (II), respectively, during the LAMP process. **b.** Similar to **a.**, hysteresis curves are shown, but for magnetic fields along *x*. The orange and azure lines represent the hysteresis curves obtained from the programmed area with $\mathbf{H}_w$ along +*x* (III) and along -*x* (IV), respectively, during the LAMP process. **c.-d.** The terahertz waveforms under the zero-field condition obtained from the programmed areas with $\mathbf{H}_w$ applied along different directions in the LAMP process. The solid lines depict the $E_x$ components, while the dashed lines represent the $E_y$ components. The signed amplitudes of the waveforms are labeled by the solid triangles. **e.** The Kerr microscopy images of the magnetic domains programmed under different writing laser powers ($P_w$). **f.** The writing resolution in diameter ($D_w$) as a function of $P_w$ obtained from the Kerr microscopic measurements. The red solid line represents the fit to the empirical model.

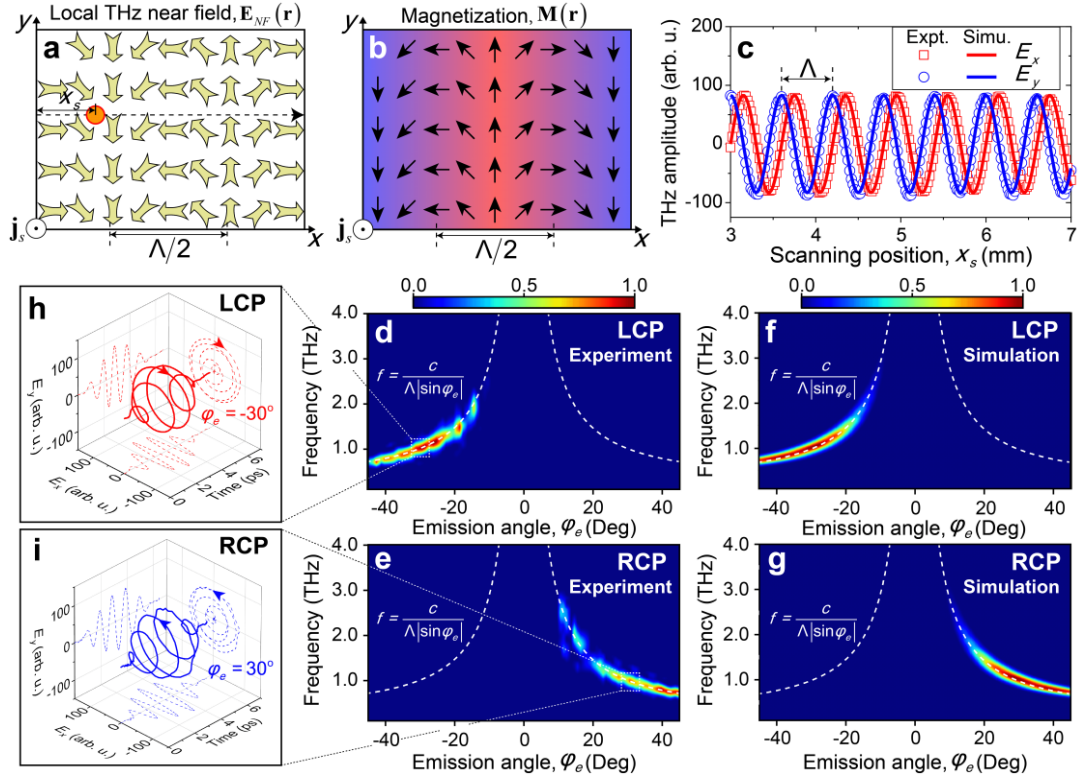

**Figure 3. Programmed terahertz emitter with spatially separated circular polarizations.**

**a.** The distribution of local terahertz NF $\mathbf{E}_{NF}(\mathbf{r})$ obtained from the design strategy. The trajectory for the focused-beam excitation measurement is illustrated with a scanning position of $x_s$. **b.** The magnetization pattern $\mathbf{M}(\mathbf{r})$ obtained from $\mathbf{E}_{NF}(\mathbf{r})$ in **a**. **c.** The symbols represent the signed amplitudes of the $E_x$ and $E_y$ fields obtained from the focused-beam excitation measurements as a function of the scanning position, $x_s$. The solid lines depict results obtained from the wave-propagation simulations with a focused excitation laser spot and the trajectory shown in **a**. **d.-e.** The spatio-spectral distribution of the experimentally measured terahertz waves with LCP and RCP, respectively, under the zero-field condition. The dashed lines represent the spatial chirp of the terahertz waves. **f.-g.** Similar to **d** and **e**, simulation results obtained from the wave-propagation simulations for the LCP and RCP components. **h.-i.** The three-dimensional temporal profiles of the terahertz waveforms emitted at $\varphi_e=+30°$ and $-30°$, respectively.

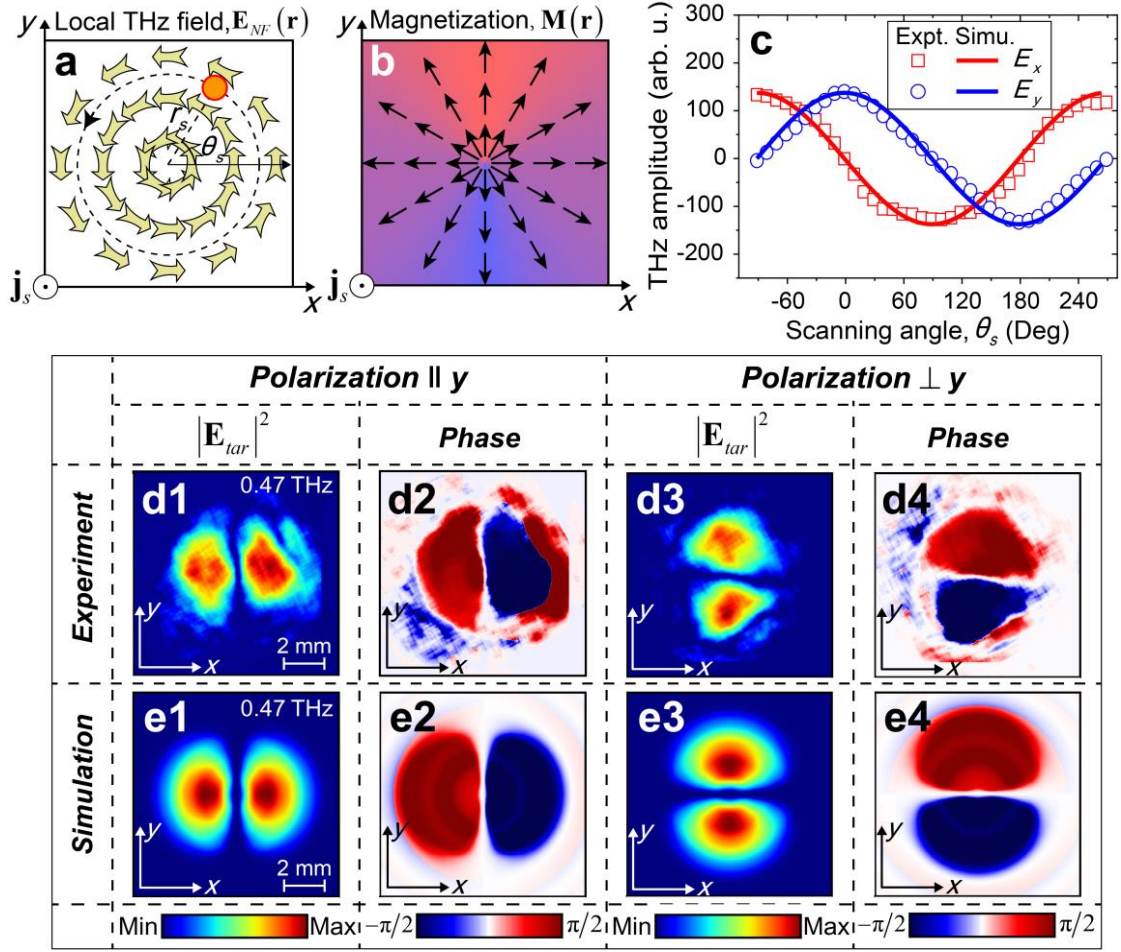

**Figure 4. Programmed terahertz emitter with azimuthal polarization. a.** The distribution of local terahertz NF $\mathbf{E}_{NF}(\mathbf{r})$ obtained from the design strategy. The trajectory for the focused-beam excitation measurement is illustrated with a radius of $r_s$ and a scanning angle of $\theta_s$. **b.** The magnetization pattern $\mathbf{M}(\mathbf{r})$ obtained from $\mathbf{E}_{NF}(\mathbf{r})$ in **a**. **c.** The symbols represent the results from the focused-beam excitation measurements as a function of the scanning angle, $\theta_s$. The solid lines depict results from the wave-propagation simulations with a focused excitation laser spot and the trajectory shown in **a**. **d1.-d4.** The experimentally measured intensity ($|\mathbf{E}_{tar}|^2$) and phase profiles of the emitted terahertz waves at a frequency of 0.47 THz for polarizations parallel to $y$ ($\parallel y$) and perpendicular to $y$ ($\perp y$). **e1.-e4.** Similar to **d**, simulation results obtained from the wave-propagation simulations.

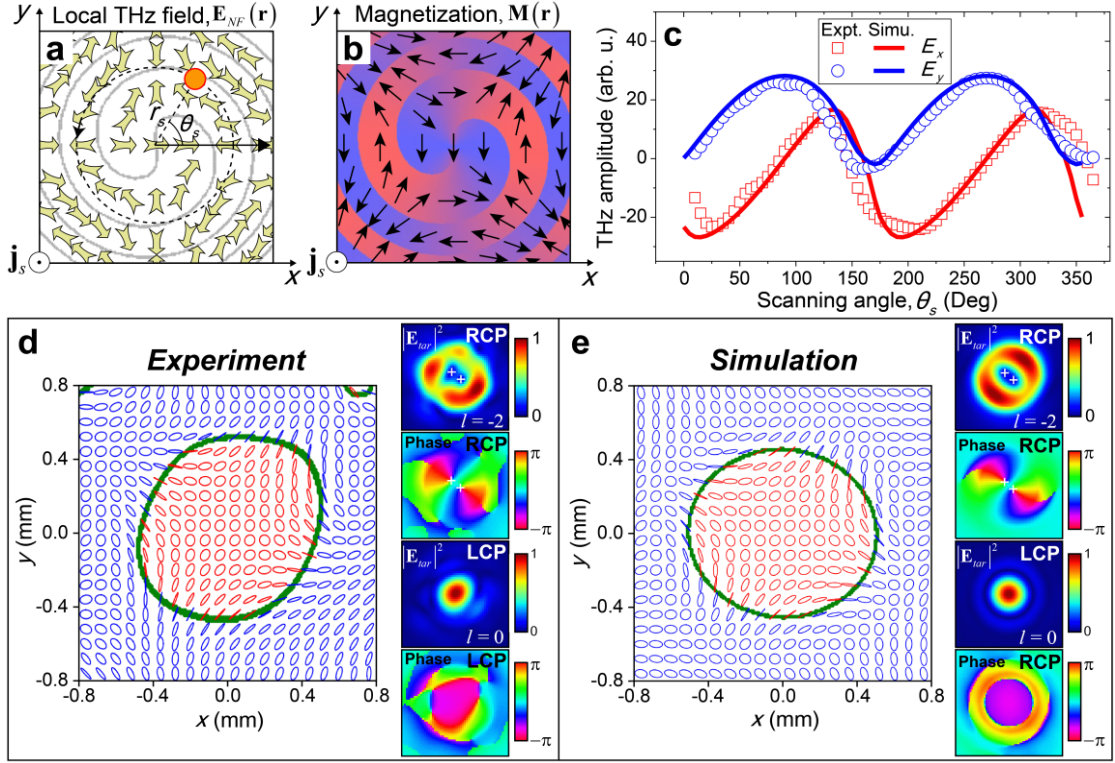

**Figure 5. Programmed terahertz emitter for the full Poincaré beam. a.** The distribution of local terahertz NF $\mathbf{E}_{NF}(\mathbf{r})$ obtained from the design strategy. The trajectory for the focused-beam excitation measurement is illustrated with a radius of $r_s$ and a scanning angle of $\theta_s$. **b.** The magnetization pattern $\mathbf{M}(\mathbf{r})$ obtained from $\mathbf{E}_{NF}(\mathbf{r})$ in **a**. **c.** The symbols represent the results from the focused-beam excitation measurements as a function of the scanning angle, $\theta_s$. The solid lines depict results from the wave-propagation simulations with a focused excitation laser spot and the trajectory shown in **a**. **d.** The experimental distribution of polarization ellipses across the beam cross-section of the FP beam. The red-colored ellipses correspond to LCP, while the blue ones represent RCP. <u>Right panel:</u> The experimentally measured intensity $|\mathbf{E}_{NF}|^2$ and phase profiles of the RCP and LCP components of the FP beam. **e.** The distribution of polarization ellipses obtained from the wave-propagation simulations. <u>Right panel:</u> The simulation results of the intensity and phase profiles of the RCP and LCP components.